\documentstyle[12pt,epsf,aps,floats]{revtex}
\begin{document}

\tighten        


\title{{\vskip -2cm}\null\hfill hep-ph/9908079\\
{\vskip 1.0cm} Relativistic Modification of the Gamow Factor }

\vspace{0.1in} \author{Jin-Hee Yoon${}^{1,2}$ and Cheuk-Yin
Wong${}^2$}

\vskip 0.5cm

\address{ ${}^1$Dept. of Physics, Inha University, Inchon, South Korea \\ 
${}^2$Physics Division, Oak Ridge National Laboratory, Oak Ridge,
Tennessee, U.S.A.  }

\date{\today}
\maketitle

\vspace{0.1in}
\begin{abstract}

In processes involving Coulomb-type initial- and final-state
interactions, the Gamow factor has been traditionally used to take
into account these additional interactions.  The Gamow factor needs to
be modified when the magnitude of the effective coupling constant
increases or when the velocity increases. For the production of a pair
of particles under their mutual Coulomb-type interaction, we obtain
the modification of the Gamow factor in terms of the overlap of the
Feynman amplitude with the relativistic wave function of the two
particles.  As a first example, we study the modification of the Gamow
factor for the production of two bosons.  The modification is
substantial when the coupling constant is large.

\end{abstract}



\narrowtext

\section{Introduction} Initial- and final- state interactions are
important in many branches of theoretical physics involving the
reaction or the production of particles.  They have a great influence
on the reaction rates or the production cross sections
\cite{Bet56,Gam28,Som39,Bar80,Gus88,Fad88,Fad90,Bro95,Cha95a,Won96d,Won97}.
These initial- and final-state interactions lead to a large
enhancement of the cross section if the particles are subject to a
strong attractive interaction; they can lead to a large suppression
under a strong repulsive interaction.  We shall use the term ``the
$K$-factor'' to label the ratio of the cross section with the
interaction to the corresponding quantity without the interaction.

As is well known, for interactions such as the electric-Coulomb and
color-Coulomb interaction $V(r)=-\alpha/r$ in non-relativistic
physics, the effect of the initial- or final-state interactions leads
to a $K$-factor given by the Gamow-Sommerfeld factor
\cite{Gam28,Som39}. The Gamow-Sommerfeld factor (or simply called, the
Gamow factor) is given explicitly by
\begin{eqnarray}
\label{eq:80}
G (\eta)
= 
{ 2 \pi \eta \over 1- e^{-2 \pi \eta}  },
\end{eqnarray}
where
\begin{eqnarray}
\label{eq:xi}
\eta={ \alpha \over v }.
\end{eqnarray}
The magnitude of the relative velocity $v$ is the ratio of the
asymptotic momentum $p$ to the energy $\epsilon_w$ in the
relative coordinate system (see Eqs.\ (\ref{eq:ew}) and
(\ref{eq:pin}) below).  Following Todorov \cite{Tod71}, Crater and Van
Alstine \cite{Cra83}, and Eqs.\ (21.13a)-(21.13c) of Crater $et~al.$
\cite{Cra92}, the relative velocity for the particles $a$ and $b$ 
is related to their center-of-mass energy $\sqrt{s}$ by
\begin{eqnarray}
\label{eq:v}
v= {(s^2 - 4s m^2)^{1/2} \over s - 2 m^2 }.
\end{eqnarray}
This gives $v \sim 2\sqrt{1-4m^2/s}$ when $\sqrt{s} \sim 2m$ and
$v\rightarrow 1$ when $s \rightarrow \infty$.  This Gamow factor has
been used to study initial- and final-state interactions in reaction
processes. 

There are physical processes in which the coupling constant of the
interaction between the particles can be quite large and the use of
the Gamow factor to correct the initial-state and final-state
interactions may not be adequate.  For example, in the annihilation or
the production of $q \bar q$ pairs, the interaction arising from the
exchange of a gluon leads to a color-Coulomb interaction with a
coupling constant $\alpha$ about 0.2-0.4, depending on the
renormalization scale of the reaction process.  Another example of
strong coupling occurs in the case of a negatively charged particle in
a nucleus with a large $Z$ number.  Such a large coupling constant
will also lead to a modification of the Gamow factor as there are
higher-order effects of the potential which are important when the
coupling constant becomes large.  One can mention, for example, the
well-known case of the ``Landau fall'' which is the relativistic
nonperturbative collapse of the wave function for an attractive
Coulomb-type potential when the coupling constant exceeds a certain
limit \cite{Lan58}.  Furthermore, although the effect of the
interaction is very large for low relative velocities, it is useful to
see how the effect varies as the velocity increases.

While one sees the need to use the relativistic formalism to study the
case with high relative velocities, one may wonder what special cases
can be of interest to use a relativistic formalism for the case of low
relative velocities.  By the term ``the relative velocity'', we
usually refer to the relative velocity between the particles in the
asymptotic region of $r \rightarrow \infty$ where there is no
interaction.  However, when there is a strongly attractive interaction,
the actual relative velocity depends on the spatial location.  One can
envisage that if the coupling constant is large, the motion of the two
particles at small distances can become relativistic, even though the
relative velocity at $r \rightarrow \infty$ is small.  Hence, it is
necessary to use the relativistic formalism to study the effects of
the mutual interaction with large coupling constants, even for the
case of low asymptotic relative velocities at $r\rightarrow \infty$.

The $K$-factor for the Coulomb potential can be studied by examining
the two-body wave function in the Klein-Gordon or the Dirac equation
involving a Coulomb potential.  Compared with the non-relativistic
Schr\" odinger equation involving the Coulomb potential, there is an
additional effective attractive potential, $-|V(r)|^2/2m_w$ (see
Eq. (\ref{eq:KG}) below), which leads to a non-trivial behavior when
the coupling constant becomes large.  In the case of fermions under
the Coulomb interaction, there are further modifications associated
with additional spin-dependent potential terms.

The question of initial- and final-state interactions is also related
to the question of the decay and the production of bound states when
the interactions lead to the formation of bound states
\cite{Pes95,Cra91}.  Previously, the decay of the bound positronium
${}^1S_0$ state into two photons has been studied in the relativistic
formalism by Crater \cite{Cra91}.  In our present study, we are
interested mainly in the case of two particles in the continuum.  We
wish to find out how the mutual interaction may affect their reaction
or production rates.

The standard method of calculating the $K$-factor is by evaluating the
absolute square of the wave function at the origin.  Such a method
breaks down for the relativistic case as the wave function is infinite
at the origin \cite{Cranote}.  The proper method to obtain the
$K$-factor is by taking the overlap of the relativistic wave function
with the Feynman amplitude.  As an illustration, we shall try out the
method for the production of a pair of scalar particles interacting
with a Coulomb-type final-state interaction.

\section{The $K$-factor}

The effects of the final- and initial-state interaction depend on the
physical process.  Here we shall be interested in the class of
processes involving the reaction or the production of a pair of
particles $a$ and $b$, subject to the mutual interaction between $a$ and $b$.
For definiteness, we shall study the production process, as the
$K$-factor is the same for production or reaction. 

The simplest description of such a process is in terms of the
perturbation theory which gives the amplitude for the production of
this pair of particles $a$ and $b$.  The state $\Phi_{a b}$ of the
$ab$ pair after the reaction $x + y \rightarrow a + b$ is represented
by the state vector
\begin{eqnarray}
\label{eq:Phi}
|\Phi_{ab} \rangle =
{\cal M}(xy
\rightarrow a(P/2+q)b(P/2-q))|a(P/2+q)b(P/2-q)\rangle
\end{eqnarray}
where ${\cal M}(xy \rightarrow ab)$ is the Feynman amplitude for the
$x + y \rightarrow a + b$ process.  For the two-particle system
$ab$, we define the center-of-mass momentum $P=a+b$ and the
relative momentum $q=(a-b)/2$.

On the other hand, under their mutual interaction which can be
represented by a two-body potential $V(r)$ between $a$ and $b$, we can
describe an $ab$ pair with a center-of-mass momentum $P$ as
\begin{eqnarray}
\label{eq:Psi}
|\Psi_V \rangle= {\tilde \psi}(q)|P\rangle.
\end{eqnarray}

The probability amplitude for the production of an $ab$ pair under
their mutual interaction is obtained by taking the overlap of the
amplitude in (\ref{eq:Phi}) with the wave function in (\ref{eq:Psi}).
The overlap is the simplest in the $ab$ center-of-mass system where
$P=(\sqrt{s},{\bf 0})$ and $q=(0,{\bf q})$ and the (unnromalized)
probability amplitude is \cite{Pes95,Cra91,Won99}
\vskip -0.4cm 
\begin{eqnarray}
\label{eq:un}
\langle\Psi_V |\Phi_{ab}\rangle = \int {d^3{\bf q} \over (2\pi)^3}
{\tilde \psi}({\bf q}) {\cal M}(xy \rightarrow a(\bbox{q})
b(\bbox{-q})). 
\end{eqnarray} 
\vskip -.2cm 
\noindent 
The $K$-factor for the occurrence of $ab$ in the state
$\Psi_V$ is then given by 
\begin{eqnarray} \label{eq:rat} K \equiv {
|\langle \Psi_V |\Phi_{ab} \rangle|^2 \over
  	   |\langle \Psi_0 |\Phi_{ab} \rangle|^2  }
={{\hbox{ (Production~cross~section~with~final-state~interaction)}} \over
{\hbox{(Production~cross~section~without~final-state~interaction)}}} 
\end{eqnarray}
where $|\Psi_0\rangle$ is the state of the $ab$ pair without their
mutual interaction.  We have used the unnormalized amplitude in Eq.\
(\ref{eq:un}) as any normalizsation constant will cancel out in the
definition of the $K$-factor in Eq.\ (\ref{eq:rat}).  The result of
the cross section calculated using the simple first-order diagram can
be corrected to include the effects of the final-state interaction by
multiplying with the $K$-factor:
\begin{eqnarray}
\label{eq:rat1}
{ {\hbox{Production~cross~section}}
\choose {\hbox{with~final-state~interaction}} }
= K \times {{\hbox{Production~cross~section}}
\choose {\hbox{without~final-state~interaction}}}.
\end{eqnarray}

\section{Klein-Gordon Equation for the Coulomb Interaction}

In this first study, in order to illustrate the main features of the
effect and to avoid complications brought on by the spinor algebra, we
shall carry out the procedures outlined above for the production of
two scalar particles.

We need to separate out the center-of-mass motion and the relative
motion for these two particles.  Consider first the case without a
mutual interaction as in the region where the two particles are far
apart.  The two particles have 4-momenta $p_1$ and $p_2$ and rest
masses $m_1$ and $m_2$.  We introduce the total momentum $P$
\cite{Cra83,Cra92}
\begin{eqnarray}
P=p_1+p_2
\end{eqnarray}
and the relative momentum $q$
\begin{eqnarray}
q=w_1 p_1 - w_2 p_2  
\end{eqnarray}
where
$$w_{1}={s - m_2^2 + m_1^2 \over 2 {s} }$$ 
and
$$w_{2}={s - m_1^2 + m_2^2 \over 2 {s} }$$
with $s=P^2$. We have the following
identity
\begin{eqnarray}
p_1^2-m_1^2+
p_2^2-m_2^2={(w_1^2 + w_2^2 )} P^2 + 2 q^2 - m_1^2 - m_2^2
=0.
\end{eqnarray}
We can choose to work in the center-of-mass system in which
$P=(\sqrt{s}, \bbox{0})$ and $q=(0,\bbox{q})$.  The above equation can be
written in terms of an effective energy $\epsilon_w$, and a
generalized reduced mass $m_w$ as
\begin{eqnarray}
\label{eq:NI}
\epsilon_w^2-\bbox{q}^2- m_w^2 = 0,
\end{eqnarray}
where
\begin{eqnarray}
\label{eq:ew}
\epsilon_w={s - m_1^2 - m_2^2 \over 2 \sqrt{s}} 
\end{eqnarray}
and
\begin{eqnarray}
m_w={m_1 m_2 \over \sqrt{s}}.
\end{eqnarray}

Next, we study the system of masses $m_1=m_2=m$ interacting with a mutual
Coulomb-type interaction
\begin{eqnarray}
V(r)=- {\alpha \over r}.
\end{eqnarray}
The equation of motion can be obtained from Eq. (\ref{eq:NI}) by the
canonical method of replacing $\epsilon_w$ with $\epsilon_w - V(r)$.
The Klein-Gordon equation for the two-particle system under a mutual
Coulomb-type interaction $V(r)$ is
\begin{eqnarray}
\label{eq:KG}
\biggl \{ [\epsilon_w - V (r) ]^2 - \bbox{q}^2 - m_w^2 \biggr \} 
\psi(\bbox{r}) = 0.
\end{eqnarray}
Writing
$\psi(\bbox{r})= R_{nl}(r) Y_{lm}(\theta,\phi)$ in (\ref{eq:KG}), the
equation for $R_{nl}(r)$ is
\begin{eqnarray}
\label{eq:psieq}
\biggl [ {d \over dz^2} + { 2 \over z} { d \over dz} - { l (l+1) \over
z^2 } + {2 \eta \over z} + {\alpha^2 \over z^2 } + 1 \biggr ]
R_{nl}(r)= 0,
\end{eqnarray}
where $z=p r$, $p$ is the asymptotic momentum at
$r\rightarrow \infty$ given by
\begin{eqnarray}
\label{eq:pin}
p= \sqrt{\epsilon_w^2 - m_w^2}.
\end{eqnarray}
The wave function $R_{nl}(r)$ can be represented by the dimensionless
variable $z=p r$ and is characterized by two dimensionless parameters:
$\eta={\alpha / v}$ and $\alpha^2$, where $v={p / \epsilon_w}$ is
given by Eq. (\ref{eq:v}).

The solution of Eq.\ (\ref{eq:psieq}) is
\begin{eqnarray}
\label{eq:wf}
R_{nl}(r)= {\left | \Gamma (a) \right| \over \Gamma (b)} e^{\pi\eta /
2} (2iz)^{\mu - 1/2} e^{-iz} {}_1F_1 (a,b,2iz)
\end{eqnarray}
where 
\begin{eqnarray}
a= \mu + {1 \over 2} + i \eta,
\end{eqnarray}
\begin{eqnarray}
b=2\mu+1,
\end{eqnarray}
\begin{eqnarray}
\mu=\sqrt{ ( l+{ 1 \over 2})^2 - \alpha^2 },
\end{eqnarray}
${}_1F_1$ is the confluent hypergeometric function, and the
normalization constant has been determined by using the boundary
condition that at $r \rightarrow \infty$, $R_{nl}(r) \rightarrow \sin
(p r + \delta_l)/p r $ with the Coulomb phase shift $\delta_l$.  For
the S-state, the critical value of $\alpha$ is 1/2.

\section{The Feynman Amplitude and the Overlap with the Coulomb Wave Function}

To obtain the effect of the final-state interaction between $a$ and
$b$ produced in a pair-production process, we consider the production
of the pair of bosons from the fusion of two photons.  Because the
relevant factors associated with the mode of production will be
cancelled out at the end in Eq.\ (\ref{eq:rat}), the results of the
$K$-factor depend only on the final-state interaction and is the same
for a similar mode of production of the pair of bosons.

\vspace*{4.8cm}
\epsfxsize=300pt
\includegraphics{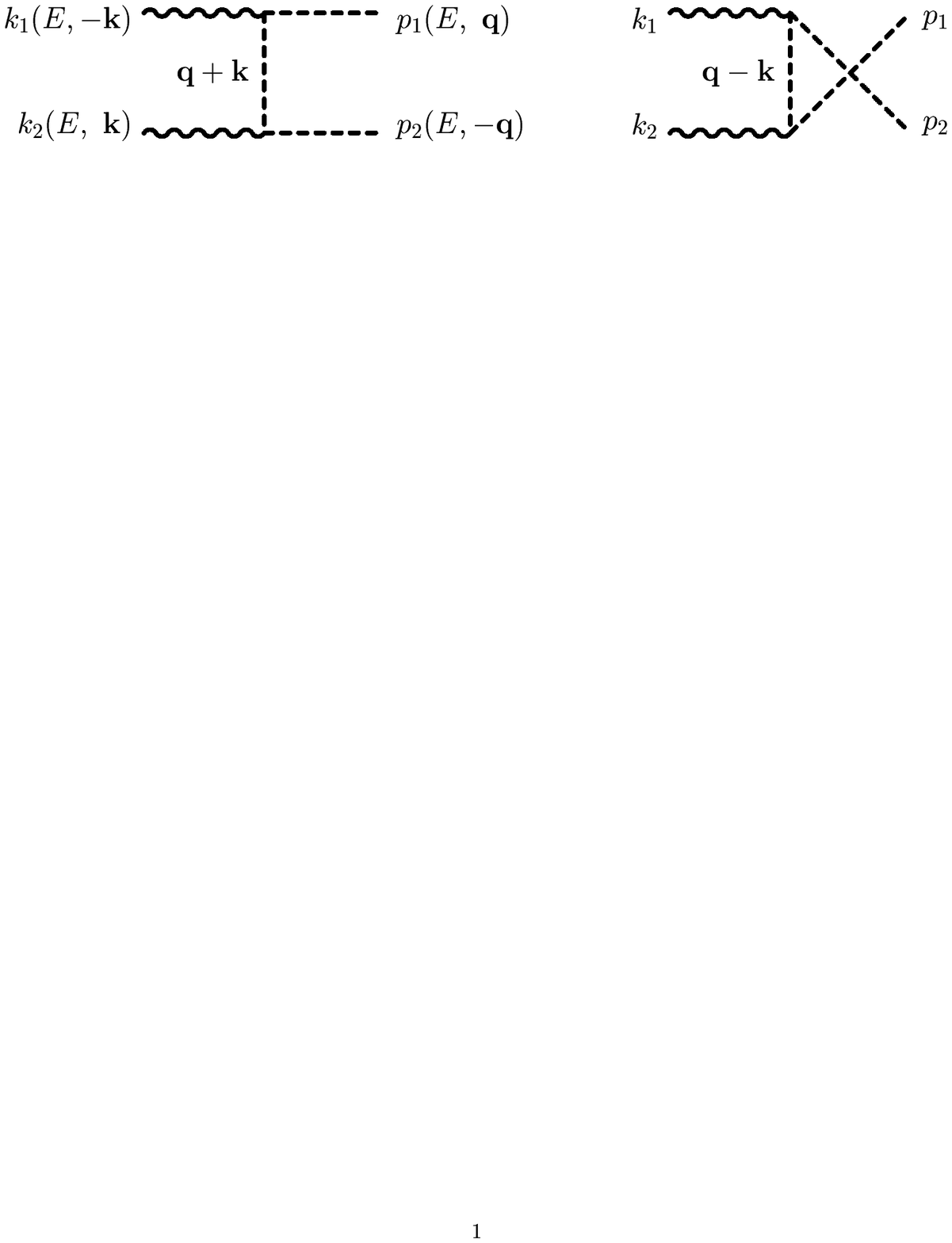}
\vspace*{-1.8cm}\hspace*{3cm}
\begin{minipage}[t]{14cm}
\noindent {\bf Fig.\ 1}.  {Feynman diagrams included in the calculation.}
\end{minipage}
\vskip 4truemm
\noindent 

The diagrams we include are shown in Fig.\ 1 which give the amplitude
\begin{eqnarray}
\label{eq:amp}
-i {\cal M}(xy \rightarrow a(P/2+q)b(P/2-q)) = i e^2
\biggl [ - {(2\bbox{q} + \bbox{k})\cdot \ \bbox{\epsilon}_1
~\bbox{k}\cdot \bbox{\epsilon}_2 \over (\bbox{q}+\bbox{k})^2+ m^2 } +
{ (2\bbox{q} - \bbox{k})\cdot \bbox{\epsilon}_2 ~\bbox{k}\cdot {\bbox
\epsilon}_1 \over (\bbox{q}-\bbox{k})^2+ m^2 } \biggr ]
\end{eqnarray}
where $\bbox{k}$ is the momentum of the photon, $\bbox{q}$ is the
momentum of one of the bosons, and $\epsilon_i$ is the polarization
vector of the $i$-th photon.  The overlap of the wave function with
the Feynman amplitude is then
\begin{eqnarray}
\langle{\Psi_V}|\Phi_{ab} \rangle
= i e^2 \int {d\bbox{q} \over (2\pi)^3} 
{\tilde \psi} (\bbox{q}) 
\biggl [ 
-  {(2\bbox{q} + \bbox{k})\cdot
{\bbox\epsilon}_1 ~\bbox{k}\cdot {\bbox \epsilon}_2
\over (\bbox{q}+\bbox{k})^2+ m^2 }
+  {(2\bbox{q} - \bbox{k})\cdot
{\bbox\epsilon}_2 ~\bbox{k}\cdot {\bbox \epsilon}_1
\over (\bbox{q}-\bbox{k})^2+ m^2 }
\biggr ] .
\end{eqnarray}
As the Coulomb wave function of Eq.\ (\ref{eq:wf}) is given in the
configuration space, it is useful to write the above integral in terms
of the wave function in configuration space. The latter is given by
\begin{eqnarray}
\label{eq:four}
\psi(\bbox{r}) = \int { d\bbox{q} \over (2\pi)^{3} } 
{\tilde \psi} (\bbox{q}) e^{-i \bbox{q}\cdot \bbox{r}}.
\end{eqnarray}

In conventional applications, one expands the Feynman amplitude
(\ref{eq:amp}) in powers of $\bbox{q}$ and keeps only the lowest order
$q$-independent term ${\cal M}_0$:
\begin{eqnarray}
{\cal M} \approx {\cal M}_0 + O(|\bbox{q}|).
\end{eqnarray}
In this approximation of taking only the leading term, Eqs.\
(\ref{eq:rat}) and (\ref{eq:four}) then give the usual $K$-factor as
the absolute square of the wave function $\psi(r)$ at the origin
\begin{eqnarray}
K=|\psi(r=0)|^2 .
\end{eqnarray}
However, such an approximation cannot be applied to our case with the
relativistic wave function as the wave function, Eq.\ (\ref{eq:wf}), is
infinite at the origin.  To avoid this singular behavior, the full
Feynman amplitude is needed to evaluate the overlap integral and the
$K$-factor in Eqs.\ (\ref{eq:amp}) and (\ref{eq:rat}). 

In terms of the spatial wave function, the overlap integral
(\ref{eq:amp}) is
\begin{eqnarray}
\langle{\Psi_V}|\Phi_{ab} \rangle
= e^2  \int d\bbox{r} \psi(\bbox{r}) 
\Biggl [& & e^{-i\bbox{k}\cdot \bbox{r}}
\{ 2 {\hat{\bbox{\epsilon}} }_1\cdot {\nabla \over i}
- {\hat{\bbox{\epsilon}}}_1\cdot \bbox{k}\} 
{e^{-mr} \over 4 \pi r } 
{\hat{\bbox{\epsilon}}}_2\cdot \bbox{k} 
\nonumber\\
&-& 
e^{i\bbox{k}\cdot \bbox{r}} 
\{ 2 {\hat{\bbox{\epsilon}} }_2\cdot {\nabla \over i}
+ {\hat{\bbox{\epsilon}} }_2\cdot \bbox{k}\} 
{e^{-mr} \over 4 \pi r } 
{\hat{\bbox{\epsilon}}}_1\cdot \bbox{k} 
\Biggr ].
\end{eqnarray}

We shall specialize to the S-wave case with $l=0$.  Using the wave
function of Eq.\ (\ref{eq:wf}), we carry out the above integration 
and obtain
\begin{eqnarray}
\label{eq:mfi}
\langle{\Psi_V}&|&\Phi_{ab} \rangle
= 2 e^2 ({\hat{\bbox{\epsilon}}}_1\cdot \bbox{k})
        ({\hat{\bbox{\epsilon}}}_2\cdot \bbox{k})
{|\Gamma(a)| \over \Gamma(b)}
e^{\pi \eta /2}
\sum_{n=0}^\infty
{ (a)_n  \Gamma ({3 \over 2 }+ \mu + n) \over (b)_n~ n! } 
 \left ({  2i p \over \sqrt{\delta^2 + k^2}} \right )^{n+\mu-{1 \over2} }
{  1 \over {\delta^2 + k^2}}
\nonumber \\
&\times& \Biggl [ {2 \over 3} F({3 \over 4} + {\mu+n \over 2}, {5\over 4}
- {\mu + n \over 2}; {5\over 2} ; \xi^2)
+ {2m \over 3} { {3\over 2 } +\mu +n \over \sqrt{\delta^2 + k^2}}
 F({5 \over 4} + {\mu+n \over 2}, {3\over 4} - {\mu + n \over 2}; 
{5\over 2} ; \xi^2)
\nonumber\\
& &
- F({3 \over 4} + {\mu+n \over 2}, {1\over 4} - {\mu + n \over 2};
{3\over 2} ; \xi^2) \Biggr ],
\end{eqnarray}
where 
\begin{eqnarray}
\delta=m+ip,
\end{eqnarray}
\begin{eqnarray}
\xi^2={k^2\over (m+ip)^2 + k^2}.
\end{eqnarray}
In Eq.\ (\ref{eq:mfi}) $(a)_n=a (a+1)(a+2)..(a+n-1)$, with $(a)_0=1$.
The quantity $(b)_n$ is similarly defined.

\section{Results for the $K$-factor }

We introduce the complex angle variable
\begin{eqnarray}
\theta = \tan^{-1}{ k \over m+ip} = {\pi \over 4} - i {1 \over 4}
\ln { k + p \over k - p},
\end{eqnarray}
which is a relativistic measure of the relative motion between
particles $a$ and $b$.  The real part of $\theta$ is always $\pi/4$,
and the imaginary part is negative, with a magnitude that is half of
the rapidity of the produced particle in the center-of-mass system.

In terms of the angle variable, Eq.(\ref{eq:mfi}) can be transformed
as follows:
\begin{eqnarray}
\label{eq:up}
\langle{\Psi_V}|\Phi_{ab} \rangle
= 2 e^2 ({\hat{\bbox{\epsilon}}}_1\cdot \bbox{k})
        ({\hat{\bbox{\epsilon}}}_2\cdot \bbox{k})
	{|\Gamma(a)| \over \Gamma(b)}
	e^{\pi \eta /2}	 {\cal A},
\end{eqnarray}
where the factor $\cal A$ is
\begin{eqnarray}
\label{eq:A1}
{\cal A}&=& 
\sum_{n=0}^\infty
{ (a)_n   \Gamma (2\nu) \over (b)_n~ n! } 
{ \left ( {2ip \over \sqrt{\delta^2 + k^2} } \right ) ^{n+\mu-{1\over 2}} } 
{1 \over k^2 {\sin \theta} }\nonumber\\
&\times&
\Biggl [ {2 \over 2\nu-3} \left\{ {\sin (2\nu-2) \theta \over 2\nu-2} 
\cos \theta
- {\sin (2\nu-1) \theta \over 2\nu-1} \right\}
\nonumber\\
&+& {2m \over k} {2\nu \over 2\nu-2} \sin \theta
	\left\{ { \sin (2\nu-1) \theta \over 2\nu-1} \cos \theta
		- {\sin 2\nu \theta \over 2\nu}
	\right\}
- \sin^2 \theta { \sin (2\nu-1) \theta \over 2\nu-1} 
\Biggr ],
\end{eqnarray}
and $\nu=3/4 + (\mu + n)/2$.  To obtain the $K$-factor, we also need
the overlap between the Feynman amplitude and the wave function
without the final-state interaction.  By using the wave function
$\psi_0 (r) = {\sin p r}/{p r}$ for the S-state without the Coulomb
potential, we obtained the amplitude for the case without the
final-state interaction as given by
\begin{eqnarray}
\label{eq:down}
 \langle{\Psi}_0 | \Phi_{ab} \rangle
	= 2 e^2  { ({\hat{\bbox{\epsilon}}}_1\cdot \bbox{k})
        ({\hat{\bbox{\epsilon}}}_2\cdot \bbox{k}) }
	{\cal B},
\end{eqnarray}
where the factor $\cal B$ is
\begin{eqnarray}
{\cal B}=
 {\cal I}m
	\left\{ ( \cot \theta^* -2m/k)(\theta^* \cot \theta^* -1)
	\right\}/pk,
\end{eqnarray}
and $\theta^*$ is a complex conjugate of $\theta$.  Then the ratio
between the absolute squares of Eqs.(\ref{eq:up}) and (\ref{eq:down})
is the relativistic expression of the $K$-factor,
\begin{eqnarray}
\label{eq:kfinal}
K &=& \left| 
	{\Gamma(a) \over \Gamma(b)}
	e^{\pi \eta /2} 
	{{\cal A} \over {\cal B}} \right |^2.
\end{eqnarray}
We can identify the factor $|\Gamma(a)e^{\pi \eta /2} / \Gamma(b)|^2$
as closely related to the Gamow factor $G(\eta)$.  One can show that
\begin{eqnarray}
\left|
        {\Gamma(a) \over \Gamma(b)}
        e^{\pi \eta /2} \right |^2 = G(\eta) \left| {\Gamma(\mu+1/2)
\over \Gamma(2\mu+1)} \prod_{j=0}^{\infty} \left ( 1+{\mu - 1/2 \over
1+j} \right )^2 \left ( 1 + {({3\over 2} + \mu + 2 j) ( {1 \over 2} -
\mu) \over (\mu + {1 \over 2} + j)^2 + \eta^2 } \right ) \right|^2 .
\end{eqnarray} 
Therefore, the proper treatment of the dynamics of the
interacting particles leads to the modification of the Gamow factor
$G(\eta)$ of Eq.\ (\ref{eq:80}) by a factor $\kappa$ given by
\begin{equation} 
K=G(\eta)\kappa 
\end{equation} 
where 
\begin{eqnarray}
\kappa=\left| {\Gamma(\mu+1/2) \over \Gamma(2\mu+1)}
\prod_{j=0}^{\infty} \left ( 1+{\mu - 1/2 \over 1+j} \right )^2 \left
( 1 + {({3\over 2} + \mu + 2 j) ( {1 \over 2} - \mu) \over (\mu + {1
\over 2} + j)^2 + \eta^2 } \right ) \right|^2 \left | {\cal A \over
\cal B} \right |^2, \end{eqnarray} and \begin{eqnarray} \label{eq:A2}
\left | {\cal A \over \cal B} \right |^2 &=& \left| \sum_{n=0}^\infty
{ (a)_n \Gamma (2\nu) \over (b)_n ~ n!  ~ {\sin \theta}} \left ( {2 i
p \over \sqrt{\delta^2 + k^2} }\right ) ^{n+\mu-{1 \over 2}}
\right. \Biggl [ {2 \over 2\nu-3} \left\{ {\sin (2\nu-2) \theta \over
2\nu-2} \cos \theta - {\sin (2\nu-1) \theta \over 2\nu-1} \right\}
\nonumber\\ & & + {2m \over k} {2\nu \over 2\nu-2} \sin \theta \left\{
{ \sin (2\nu-1) \theta \over 2\nu-1} \cos \theta - {\sin 2\nu \theta
\over 2\nu} \right\} \left. - \sin^2 \theta { \sin (2\nu-1) \theta
\over 2\nu-1} \Biggr ] \right|^2 \nonumber\\ &/ & \left|(k/p) {\cal
I}m \left\{ (k \cot \theta^* -2m)(\theta^* \cot \theta^* -1) \right\}
\right|^2. \end{eqnarray} 
In the limit of $\alpha \rightarrow 0$ or $v\rightarrow 0$, the factor
$\kappa$ goes to 1 and is consistent with the Gamow factor.

We note that the center-of-mass energy $\sqrt{s}$ 
in units of the rest mass of
the produced particle is a function of $\eta/\alpha$:
\begin{eqnarray}
 { \sqrt{s} \over m }   = \sqrt{ 2 \left ( 1 + { {\eta/\alpha} 
\over \sqrt{ \eta^2/\alpha^2 -1}} \right )} .
\end{eqnarray}
Various other kinematic variables, such as ${k / m} = {\sqrt{s} / 2
m}$ and ${p / m} = \sqrt{ {s/ 4m^2 }-1}$ can be similarly expressed as
a function of $\eta/\alpha$.  From these relations and the relation
between the $K$-factor and $\eta$ and $\alpha$, we can study the
$K$-factor for the production of a pair of particles in a specific
kinematic configuration.

\vspace*{6.3cm} 
\epsfxsize=300pt 
\includegraphics{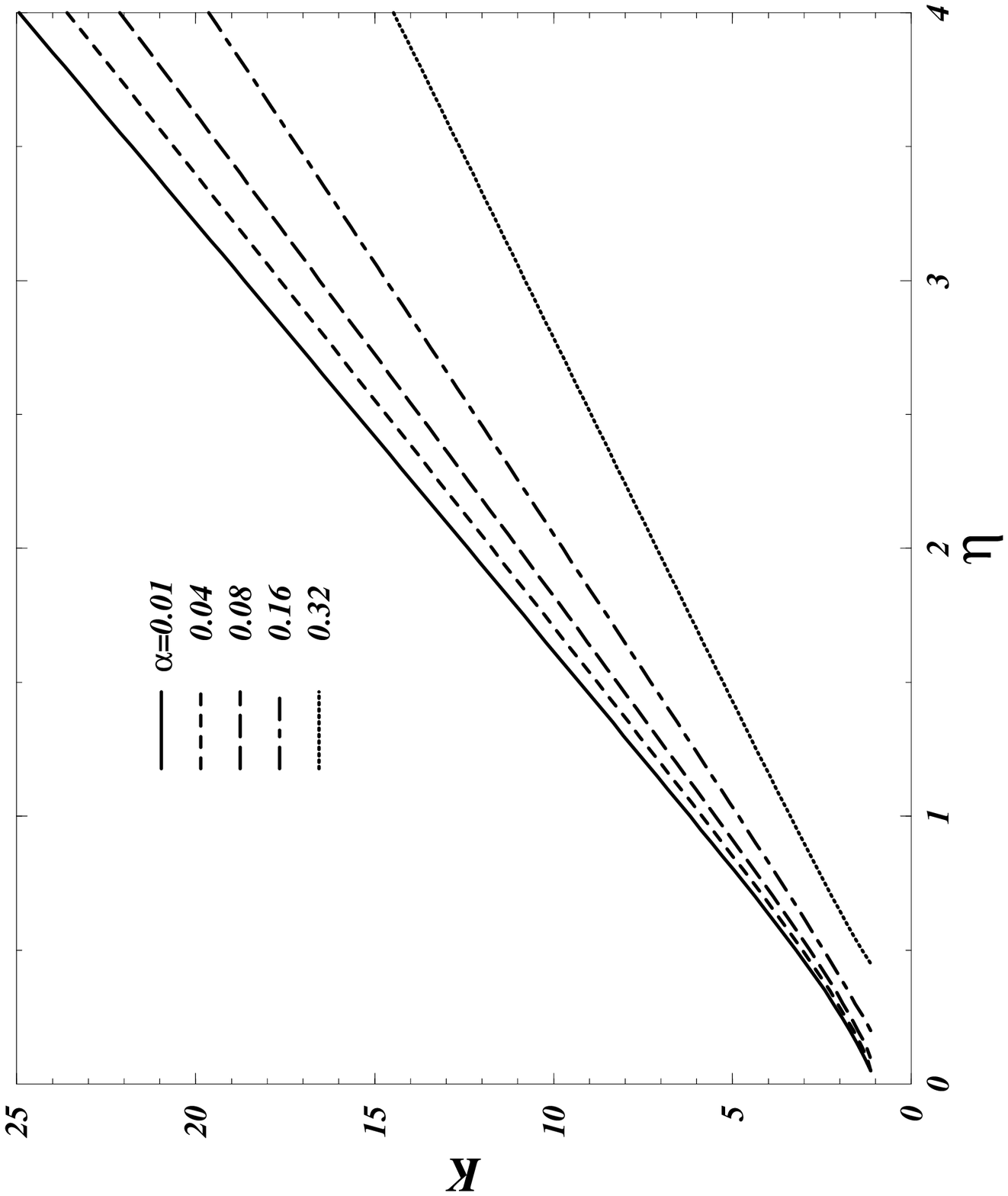} 
\vspace*{3.5cm}\hspace*{3cm}
\begin{minipage}[t]{10cm} 
\noindent {\bf Fig.\ 2}.  {The $K$-factor
versus $\eta$ for various values of $\alpha$.}
\end{minipage} 
\vskip 4truemm 
\noindent 
\vspace*{0.1cm}\noindent

We show the behavior of the $K$-factor as a function of $\eta$ in
Fig. 2 for various values of $\alpha$.  The solid curve gives the
$K$-factor for $\alpha=0.01$ and the dotted curve gives the $K$-factor
for $\alpha=0.32$.  For a fixed value of $\alpha$, the $K$-factor
decreases as $\eta$ decreases.  This is consistent with the
expectation that the effects of the final-state interaction diminish
as the velocity becomes relativistic.  The limiting value is
$K$=1 as $\eta=\alpha/v \rightarrow \alpha$.  Figure 2 also shows that
for a given value of $\eta=\alpha/v$, the $K$-factor decreases as
$\alpha$ increases.  It should be noted that the same value of $\eta$
corresponds to different velocities $v$ for different values of
$\alpha$.  To see the effect of final-state interaction as a function
of $\alpha$ for a fixed value of $v$, we plot in Fig.\ 3 the
$K$-factor as a function of $v$.  As one observes, when the velocity
is fixed, the $K$-factor increases as the coupling constant increases,
indicating a greater effect of the final-state interaction as $\alpha$
increases.  For all values of $\alpha$, the $K$-factor decreases with
$v$ and goes to unity as $v$ approaches 1.  The decrease is very rapid
for small values of $\alpha$.

\vspace*{6.3cm} 
\epsfxsize=300pt 
\includegraphics{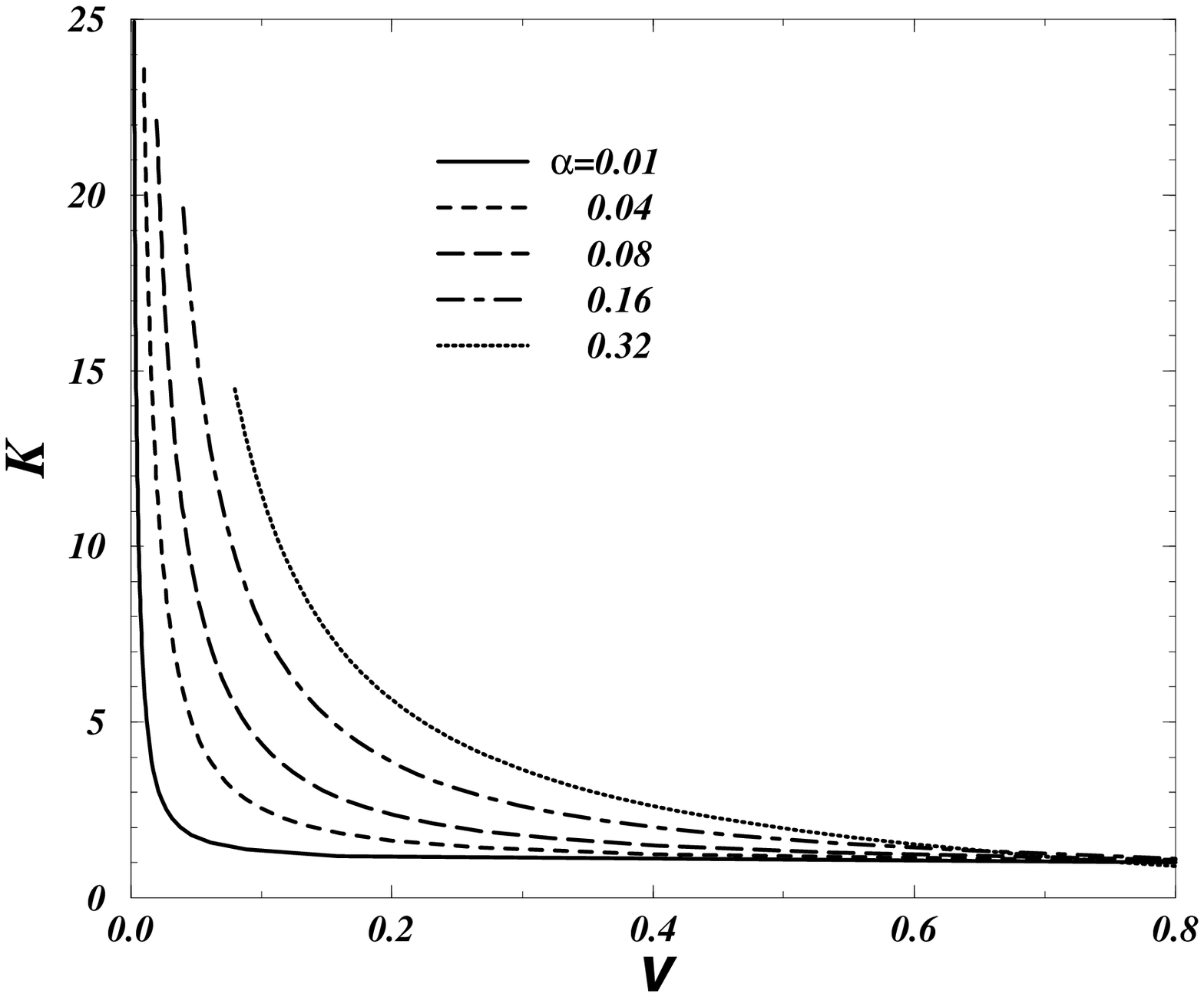} 
\vspace*{3.5cm}\hspace*{3cm}
\begin{minipage}[t]{10cm} 
\noindent {\bf Fig.\ 3}.  {The $K$-factor
versus the velocity $v$ for various values of $\alpha$.}
\end{minipage} 
\vskip 4truemm 
\noindent 
\vspace*{0.5cm}

It is of interest to see how the $K$-factor obtained here is different
from the Gamow factor in non-relativistic physics.  In Fig. 4, we
showed the ratio between the $K$-factor and the Gamow factor for
various values of $\alpha$.  As we expect, the ratio is almost 1 for
weak coupling and the use of the Gamow factor is relatively safe
there. However, if we increase $\alpha$ to 0.32, the ratio decreases
significantly.  The Gamow factor overestimates the magnitude of the
final-state interaction.  It cannot be used for the case with strong
coupling.  There is an effective screening of the long-range Coulomb
interaction.  As a consequence, the enhancement due to the long-range
Coulomb-type interaction is reduced.  It can also be observed in Fig.\
4 that the ratio of $K/G(\eta)$ is a relatively slowly varying
function of $\eta$ for $\eta>1$ but drops down rapidly as $\eta$
decreases in the region of small $\eta$.

It is worth pointing out that the expansion of $\cal A$ in Eqs.\
(\ref{eq:A1}) and (\ref{eq:A2}) is given as a series in powers of
$p/\sqrt{\delta^2+k^2}$ which increases as the velocity $v$ increases.
We still obtain convergent results for $v$ up to about 0.8, but there
is a limit on using such an expansion for greater velocities where
$p/\sqrt{\delta^2+k^2}$ is too large to allow for a convergent
term-by-term summation.  A different expansion method is needed.

\vspace*{7.8cm}
\epsfxsize=300pt
\includegraphics{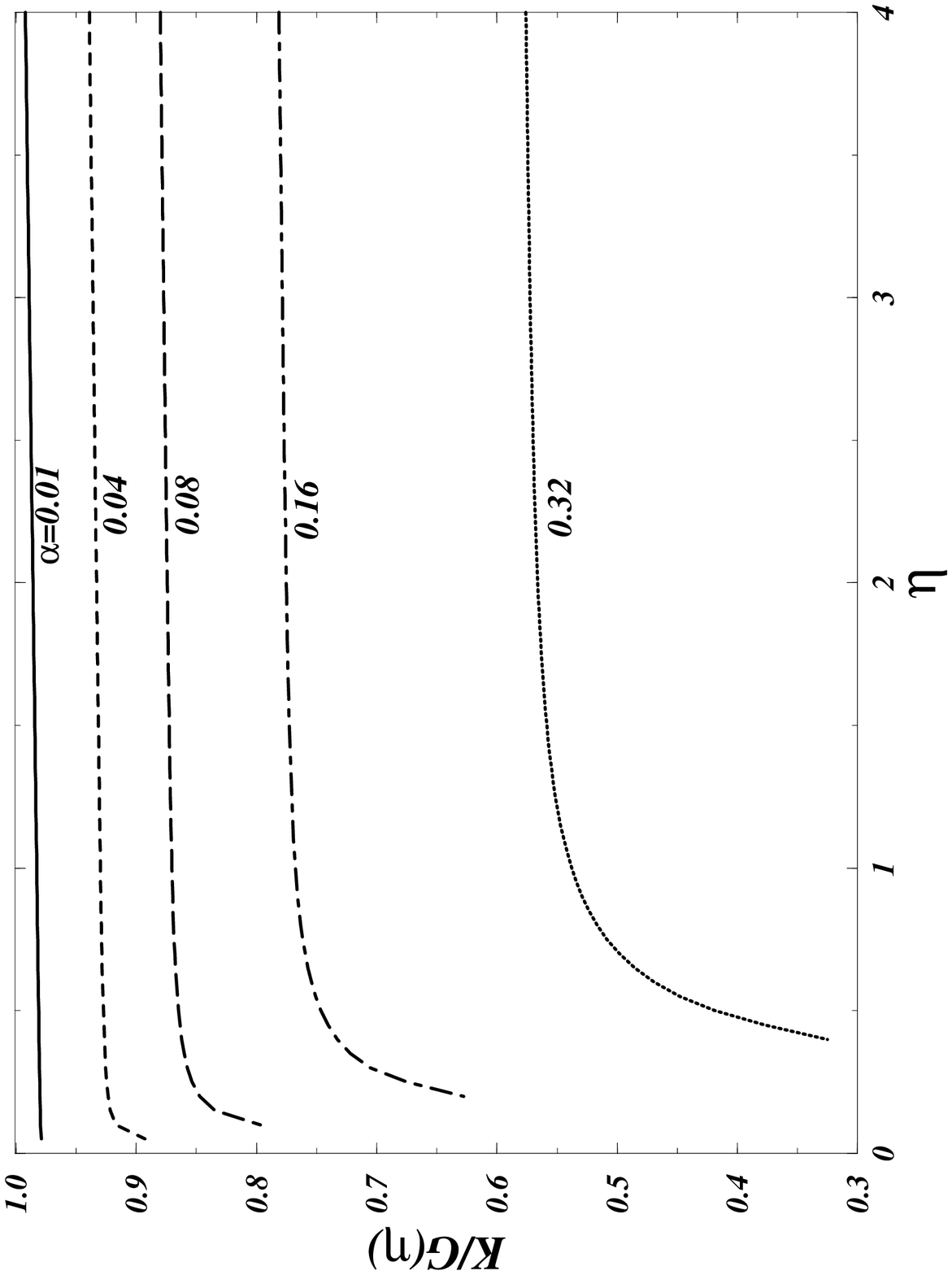}
\vspace*{2.8cm}\hspace*{2cm}
\begin{minipage}[t]{10cm}
\noindent {\bf Fig.\ 4}.  {The ratio between the $K$-factor and the Gamow
factor $G(\eta)$ 
for various values of $\alpha$.}
\end{minipage}
\vskip 4truemm
\noindent 

\section{Conclusions and Discussions}

The mutual final-state interaction between the produced particles has
an effect on their rate of production.  There will be similar effects
if the particles interact via the initial-state interaction.  The effects
are simplest to be taken into account by using the method of the
$K$-factor.  One calculates the rate for the process as though there
were no initial- or final-state interactions, using, for example, the
perturbation theory.  The additional initial- or final-state
interactions can be included by multiplying a $K$-factor as given by
Eq.\ (\ref{eq:rat1}).

For Coulomb-type interactions, the $K$-factor has been traditionally
taken to be the Gamow factor obtained as the absolute square of the
wave function at the origin of the relative coordinate.  With
relativistic Coulomb wave functions, the wave function at the origin
is infinite and the usual method is not applicable.  The $K$-factor
can be obtained as the overlap of the wave function with the Feynman
amplitude.

Our investigation of the $K$-factor for the case of the production of
a pair of scalar particles indicates that there are substantial
deviations from the Gamow factor when the strength of the coupling is
large.  In particular, the proper treatment reduces the magnitude of
the Gamow factor significantly.  The reason for this reduction is that
in the pair production, there is an effective screening of the
Coulomb-type interaction arising from the effective ``exchange'' of
one of the produced particles.

We have presented an explicit formula for the relativistic
modification of the Gamow factor for the production of a pair of
bosons.  Numerical results are also obtained to show the magnitude of
the $K$-factor.  The results of the $K$-factor can be applied to a
class of processes in which the boson particles are produced and 
interacting with a Coulomb-type interaction.

The large modification of the Gamow factor for the production of two
bosons studied here indicates the need to extend the present formalism
to study the case of two fermions or two gluons.  The application
to fermions or gluons will be useful in the problem of production or
reaction of quarks and gluons.  As a pair of quarks or gluons interact
with a strong color-Coulomb interaction with a coupling constant
$\alpha$ about 0.2-0.4, the simple results from the present study
indicate that the effects of the initial- or final-state
interaction for quark and gluon will be large, and the $K$-factor will
be substantially different from what one obtains using the Gamow
factor.  We hope to study the modification of the Gamow factor for two
produced fermions in our next investigation.

\section*{Acknowledgments}
\vspace {-0.5cm}

Tha author would like to thank Dr.  H. Crater for helpful discussions
and valuable suggestions.  This research was supported by the Division
of Nuclear Physics, U.S.D.O.E.  under Contract No. DE-AC05-96OR21400
managed by Lockheed Martin Energy Research Corp.

\end{document}